\documentclass[11pt]{article}
\usepackage{graphicx}
\title{Microscopic model for the logarithmic size effect on the Curie point in Barab\'asi-Albert networks}
\author{Agata Aleksiejuk-Fronczak}

\begin{document}
\maketitle
{\center Institute for Theoretical Physics, Cologne University, \\
D-50923 K\"oln, Euroland } {\center and} {\center Faculty of
Physics, Warsaw University of Technology,\\ Koszykowa 75,
PL-00-662 Warsaw, Poland \footnote{permanent address}}.
\bigskip
\begin{abstract}
We found that numbers of fully connected clusters in
Barab\'asi-Albert (BA) networks follow the exponential
distribution with the characteristic exponent $\kappa=2/m$. The
critical temperature for the Ising model on the BA network is
determined by the critical temperature of the largest fully
connected cluster within the network. The result explains the
logarithmic dependence of the critical temperature on the size of
the network $N$.
\end{abstract}

\bigskip
\bigskip
% new begin
During the last few years studies of random, evolving networks
have become a very popular research domain among physicists. A lot
of efforts were put into investigation of such systems in order to
recognize their structure and to analyze emerging complex
dynamics. The Barab\'asi-Albert (BA) network \cite{BA} being the
subject of this paper is probably the most studied model that
realizes many properties of real weblike systems. The most
important feature of the model is the power-law degree
distribution that is also a characteristic of many real networks.
It was shown that this feature is closely related to processes
governing system evolution. Two important ingredients of the BA
model are: continuous network growth and preferential attachment.
The network starts to grow from an initial cluster of $m$ fully
connected sites. Each new node that is added to the network
creates $m$ links that connect it to previously added nodes. The
preferential attachment means that the probability of a new link
to end up in a vertex $i$ is proportional to connectivity of this
vertex $k_{i}$.

The goal of the investigation run here was to understand the
logarithmic dependence of the Curie temperature in ferromagnetic
phase transition in BA network \cite{ising} on the system size. A
mean field theory for this behaviour has been recently found by
Bianconi \cite{Bianconi} and independently developed by
Dorogovtsev at al. \cite{Dorogov}. Here we provide another
explanation for the fact and associate it with the microscopic,
internal structure of the BA networks.

\begin{figure}[hbt]
\begin{center}
\includegraphics[scale=0.5]{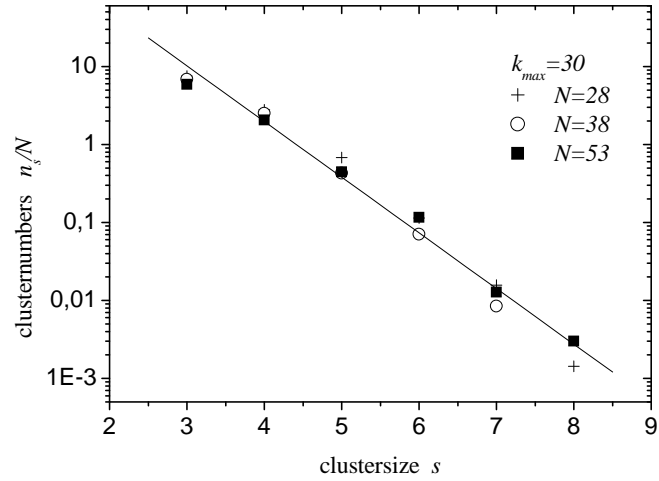}
\vskip 2cm
\includegraphics[scale=0.5]{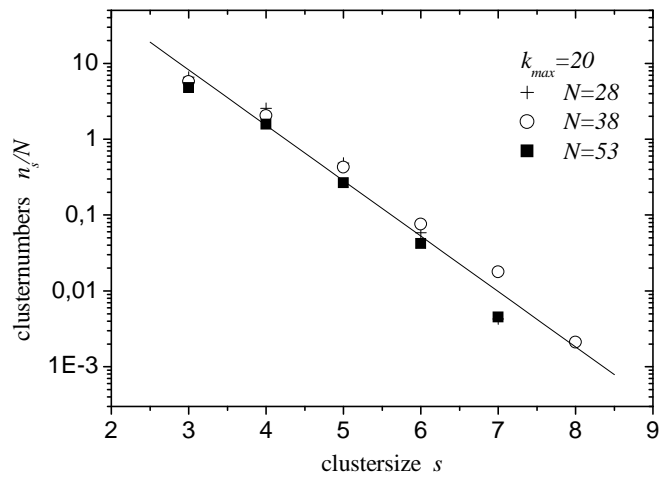}
\vskip 1cm
\end{center}
\caption{\label{fig1}Distributions of fully connected clusters,
$k_{max}=30$ (a) or $20$ (b) (averaged over $50$ networks)}
\end{figure}

We investigate the probability to occur a fully connected cluster
of size $s$ in BA networks. A fully connected $s$-cluster is
defined as a group of $s$ points, such that every point of the
group has a connection to each of the $(s-1)$ remaining points
within this group. In order to get reliable numbers $n_{s}$ of
fully connected $s$-clusters all $s$-subsets of the whole set of
network vertices should be examined. It gives $(^{N}_{s})$
possibilities for each cluster of size $s$ and altogether $2^{N}$
combinations for all cluster sizes. Such calculations would
involve unreasonable amount of computer time. To simplify the task
we noticed that each point $i$ that has $k_{i}$ nearest neighbors
may be involved in $(^{k_{i}}_{s-1})$ fully connected clusters of
size $s$. Because of computer time limitations we were forced to
restrict our studies to nodes with degree $k_{i}\leq k_{max}$
where $k_{max}=20$ or $30$ and small lattices $N=28$, $38$, $53$
(Fig.\ref{fig1}). All simulations were done for BA networks with
$m=3$. Results presented at Fig.\ref{fig1} are exact in this sense
that all possible combinations of fully connected clusters up to
size $s=k_{max}+1$ having at least one node $i$ of degree
$k_{i}\leq k_{max}$ were checked. Numerical simulations carried
out for networks of size $N$ show that the cluster numbers $n_{s}$
follow the exponential distribution:
\begin{equation}
n_{s}= A N e^{-{\rm \kappa }\ast s}. \label{e1}
\end{equation}
At the Fig.\ref{fig1} numbers of such $s$-clusters per network
node $n_{s}/N$ are fitted to (\ref{e1}) giving
\begin{eqnarray}
\ln(A)=3.20\pm0.20 \label{e1a} \\ \kappa=0.70\pm0.05 \nonumber.
\end{eqnarray}

The critical temperature for the Ising model on the fully
connected lattice that consists of $s$ sites follows the
well-known mean-field result
\begin{equation}
T_{c}=(s-1),\label{e2}
\end{equation}
where $(s-1)$ is the mean number of nearest neighbors (temperature
is measured in units of coupling constant over Boltzman constant
$J/k$). Let us assume that the critical temperature for the
largest fully connected cluster within the BA network is given by
(\ref{e2}). Since fully connected clusters in BA model are not
isolates lattices but they are involved in the whole network
structure the equation (\ref{e2}) is valid only for the largest
cluster. Putting $n_{s}=1$ and using (\ref{e1}) we get the
temperature as:
\begin{equation}
T_{c}=\frac{\ln(A)}{\kappa}-1+\frac{\ln(N)}{\kappa}. \label{e3}
\end{equation}
Taking into account the result (\ref{e1a}) for the BA model with
$m=3$ we get:
\begin{equation}
T_{c}=B\ln(N)+C, \label{e3a}
\end{equation}
where $B=1.4\pm0.1$ and $C=3.6\pm0.1$. Looking at the result
obtained by Dorogovtsev at al. \cite{Dorogov} for the whole BA
network:
\begin{equation}
T_{c}=m\ln(N)/2 \label{e4}
\end{equation}
and putting $m=3$ into (\ref{e4}) the critical temperature of the
whole network gains the form:
\begin{equation}
T_{c}=1.5\ln(N), \label{e4a}
\end{equation}
that for large $N$ fairly good agrees with (\ref{e3a}) and
(\ref{e3}) giving $\kappa=2/m$. It follows that the Curie
temperature of the whole BA network is closely related to the
Curie temperature of the largest fully connected cluster. To
destroy the magnetisation of the network one needs to destroy
ferromagnetic coupling within this cluster. Our studies are
consistent with the general conclusion of the paper \cite{Dorogov}
that the most connected vertices in random graphs with arbitrary
degree distributions induce strong ferromagnetic correlations in
their close neighborhoods at high temperatures. Such vertices
probably participate in the fully connected cluster making the
ordering phenomena of the whole network dependent on ordering of
the largest fully connected cluster.

In summary, we presented a possible microscopic explanation for
logarithmic size effect of the Curie temperature in
Barab\'asi-Albert networks. We found that the critical temperature
of the whole network is determined by the critical temperature of
the largest fully connected cluster within the network.

Acknowledgments: This project has been partially supported by DAAD
and the special grant of Warsaw University of Technology. We are
grateful to Ditrich Stauffer and Janusz Ho\l yst for helpful
discussions.

\begin{thebibliography}{99}
\bibitem{BA} A.L. Barab\'asi and R. Albert, Science {\bf 286}, 509 (1999);
R. Albert and  A.L. Barab\'asi, Rev. Mod. Phys. {\bf 74} 47 (2002).
\bibitem{ising} A. Aleksiejuk, J.A. Ho\l yst, D. Stauffer, Physica A {\bf 310} 260 (2002),cond-mat/0112312.
\bibitem{Bianconi} G. Bianconi priv. comm.
\bibitem{Dorogov} S.N. Dorogovtsev, A.V. Goltsev, J.F.F. Mendes, cond-mat/0203227.

\end{thebibliography}
\end{document}